\documentclass[times]{cpeauth}

\usepackage{setspace, amsmath, amssymb, url, lscape, subfigure, multirow, pslatex, listings, verbatim, alltt, amsfonts, wrapfig, boxedminipage, color, cite, balance}

\usepackage[ruled,linesnumbered]{algorithm2e}
\usepackage[colorlinks, citecolor=blue]{hyperref}
\usepackage{color, soul}
\usepackage{epsfig}
\usepackage{tikz}

\usepackage{xcolor}
\usepackage{leqno}
\usepackage[labelformat=parens,labelsep=quad, skip=3pt]{caption}
\usepackage [autostyle, english = american]{csquotes}
\MakeOuterQuote{"}
\newcommand{\system}{S3BD}

\newcommand{\eg}{{\it e.g., }}
\newcommand{\etal}{{\it et~al. }}
\newcommand{\ie}{{\it i.e., }}

\newcommand{\comments}[1]{}

\newlength{\boxfigwidth}

\newcommand{\boxfig}[1]{
\begin{figure}[htbp]
\begin{center}
\begin{small}
\setlength{\boxfigwidth}{3.1in}
\addtolength{\boxfigwidth}{0in}
\noindent\framebox{\quad\begin{minipage}{\boxfigwidth}
#1
\vspace{-19pt}
\end{minipage}\quad}
\end{small}
\end{center}
\end{figure}
}

\newcommand{\boxfigsecond}[1]{
\begin{figure}[htbp]
\begin{center}
\begin{small}
\setlength{\boxfigwidth}{3.1in}
\addtolength{\boxfigwidth}{0in}
\noindent\framebox{\quad\begin{minipage}{\boxfigwidth}
#1
\vspace{-47pt}
\end{minipage}\quad}
 \vspace{24pt}
\end{small}
\end{center}
\end{figure}
}

\begin{document}


\title{S3BD: Secure Semantic Search over Encrypted Big Data in the Cloud}
\author{Jason W. Woodworth\affil{1}\corrauth and Mohsen Amini Salehi\affil{2}\corrauth}

\address{\centering \affilnum{2} High Performance Cloud Computing (HPCC) Laboratory \\
\affilnum{1}\affilnum{2} School of Computing and Informatics \\
University of Louisiana at Lafayette, Louisiana, USA }

\corraddr{School of Computing and Informatics, University of Louisiana at Lafayette, Louisiana, USA. Email: jww7675@louisiana.edu, amini@louisiana.edu}
%
%

\begin{abstract}
Cloud storage is a widely utilized service for both personal and enterprise demands. However, despite its advantages, many potential users with enormous amounts of sensitive data (big data) refrain from fully utilizing the cloud storage service due to valid concerns about data privacy.  An established solution to the cloud data privacy problem is to perform encryption on the client-end. This approach, however, restricts data processing capabilities (\eg searching over the data). Accordingly, the research problem we investigate is how to enable real-time searching over the encrypted big data in the cloud. In particular, semantic search is of interest to clients dealing with big data. To address this problem, in this research, we develop a system (termed S3BD) for searching big data using cloud services without exposing any data to cloud providers. To keep real-time response on big data, S3BD proactively prunes the search space to a subset of the whole dataset. For that purpose, we propose a method to cluster the encrypted data. An abstract of each cluster is maintained on the client-end to navigate the search operation to appropriate clusters at the search time. Results of experiments, carried out on real-world big datasets, demonstrate that the search operation can be achieved in real-time and is significantly more efficient than other counterparts. In addition, a fully functional prototype of S3BD is made publicly available.

\end{abstract}


\maketitle

%
%
 
\section{Introduction}\label{sec:intro}

Cloud storage has become an inevitable solution for companies and individuals who desire to store a huge volume of data, known as big data, and relieves them from the burden of maintaining storage and processing infrastructure \cite{Zobaed2018}. However, despite the advantages cloud solutions offer, many potential clients abstain from using them due to valid concerns over data security and privacy on cloud servers~\cite{forrester:banks,javanmard,amini:secureauth} and in the data transmission process \cite{Keke1, Keke2}. For example, 73\% of banks list data privacy and confidentiality as a reason for not using cloud services, making it the most cited concern~\cite{forrester:banks}. Thus, enhancing cloud privacy and confidentiality for the users' data is of paramount importance.

Cloud storage providers commonly offer security by encrypting user data on their servers and maintaining their encryption keys. However, this approach makes the data prone to attacks, particularly, internal attackers who can have access to the encryption keys~\cite{internalattacks}. One proven solution that addresses this vulnerability is to perform the encryption on the user's end~\cite{song00}, before it is transferred to the cloud. Unfortunately, this solution limits the user's ability to interact with the data, most notably the ability to search over it. The abilities are further limited when dealing with big data where performing any possible operation on the encrypted data becomes cost- and time-prohibitive~\cite{song00}. 

Our motivation, in this research, is an organization that owns a big data scale dataset containing confidential data. One example of such an organization is a law enforcement agency with encrypted police reports and officers who need to search over the reports with their handheld devices (\eg smartphones). Users of the organization may not remember exact keywords in the documents they are looking for, or need to retrieve documents semantically related to what they are searching for. For instance, the user searches for ``burglary'' but is interested in finding documents about ``robbery'' too. As such, users require the ability of \emph{semantic search} on the encrypted big data. As the users perform the search on their handheld devices with limited processing and storage capabilities, any solution for them should not impose a major processing or storage overhead. Ideally, the users need a transparent system that enables them to only enter search queries in plain text and retrieve documents in \emph{real-time} and \emph{ranked} in order of semantic relevance. Finally, any solution should not reveal any sensitive data to internal or external attackers.

Although solutions for searching over encrypted data exist, they often do not consider the semantic meaning of the user's query. That is, they only consider the keywords entered by the user (\eg~\cite{keywordsearch:CAO}), and not the terms semantically related to the user's in the query. Other solutions do not rank documents based on their relevance to the query, imposing additional search time for the user to look through results themselves. Many solutions impose a large processing and memory overhead  (\eg~\cite{sun}), making the search service costly on cloud and possibly non-real-time. 

In this research, we offer a solution for providing semantic search over encrypted big data in real-time and with low overhead using cloud services. In our solution, each document is parsed before uploading to extract key phrases that represent the document's semantic. The key phrases are then encrypted and stored in an index structure on the cloud for search processing. Semantic information is injected into the query at search-time to be searched in the index. 

Previous similar solutions are based on using a potentially huge central index, which is fully traversed for each search query (\eg~\cite{curtmola}). Undoubtedly, these solutions do not scale for big data. A common approach to reduce the impact of a large central index is to evenly partition it into disjoint clusters (\eg~\cite{Google}), which facilitates the parallelization of searching. However, this practice is still inefficient, as much of the index content is irrelevant for any individual search query.


Another approach for reducing the impact of a large central index is to fracture the central index into topic-based clusters (also known as shards)~\cite{XuCroft}. Then, at the search time, a subset of shards are proactively chosen to be searched over. Although this solution substantially reduces the search time and required resources, the remaining problem is \emph{how to generate topic-based shards on encrypted data due to a lack of semantic information}?

More specifically, in this research, we define the problem of providing a secure cloud-based semantic search system for big data as needing to answer the following three questions:
\begin{itemize}
\item How to fracture a central encrypted index into topic-based shards without revealing semantic data to the cloud?
\item How to narrow the search operation to only shards that are relevant to the user's query, hence, increase the real-timeness of the secure semantic search operation for big data?
\item How to rank results of a search based on semantic relevance to the user's query?
\end{itemize}

In this paper, we present \textbf{S}ecure \textbf{S}emantic \textbf{S}earch over encrypted \textbf{B}ig \textbf{D}ata in the \textbf{C}loud (\system) to address the aforementioned questions. The core of S3BD is based on extracting and encrypting semantic key phrases from provided documents and clustering them into topic-based shards. To provide a real-time response to a provided search query, S3BD proactively determines the shards relevant to the query at search time and limits the search only to those shards. \system~extracts key phrases and performs data encryption only at the user-end, thus, the cloud and outside world can see nothing about the plaintext data.

In summary, \system~improves upon previous work in the literature by limiting the impact of a large encrypted central index created when using big data. Specifically, the contributions of this paper are as follows:
\begin{itemize}
\item Developing a secure, scalable, and space-efficient system for semantic searching over big data in the cloud.
\item Proposing a novel application of k-means clustering to an encrypted central search index to create topic-based clusters without using explicit semantic data.
\item Proposing a novel method for pruning a large number of shards into a small number of those most relevant to a search query through semantic comparison of the query to small samples of each shard.
\item Providing a method for ranking search results based on their semantic relevance to the query without exposing any semantic information to the cloud.
\item Evaluating and analyzing the performance, scalability, overhead, and accuracy of \system~when compared with previous works in the literature.
\end{itemize}

A prototype of \system~has been implemented and is made available to the public\footnote{The prototype can be obtained from http://hpcclab.org/products/S3BDJars.zip} used for performance evaluations. The evaluations conducted on real-world datasets demonstrate the practicality of \system~for big data.

The rest of the paper is organized as follows. Section \ref{sec:rw} reviews related works in the literature, establishing the need for our solution.  Section \ref{sec:arch} gives an overview of our proposed system architecture and explains the upload, cluster, and search processes that define \system.  Section \ref{sec:security} reviews the threat model we are working with and provides a security analysis of our solution. Section \ref{sec:eval} presents the results of our evaluations using real-world datasets.  Finally, section \ref{sec:conc} concludes the paper.
\section{Related Work}\label{sec:rw}
We provide a review on research works undertaken in the four fields most related to this work and position the contribution of our works against them. Specifically, these fields are searchable encryption, semantic searching, semantic searching over encrypted data, and clustering methods for searching.

\subsection{Searchable Encryption}
Solutions for searchable encryption (SE) are imperative for privacy preservation on the cloud.  The majority of SE solutions follow one of two main approaches, the first of which being to use cyryptographic algorithms to search the encrypted text directly.  This approach is generally chosen because it is provably secure and requires no storage overhead on the server, but solutions utilizing this method are generally slower \cite{song00}, especially when operating on large storage blocks with large files.  This approach was pioneered by Song \etal~\cite{song00}, in which each word in the document is encrypted independently and the documents are sequentially scanned while searching for tokens that match the similarly encrypted query.  Boneh \etal produced a similar system in \cite{boneh04} which utilized public key encryption to write searchable encrypted text to a server from any outside source, but could only be searched over by using a private key.  While methods following this approach are secure, they often only support equality comparison to the queries, meaning they simply return a list of files containing the query terms without ranking.

The second major approach is to utilize database and text retrieval techniques such as indexing to store selected data per document in a separate data structure from the files, making the search operation generally quicker and well adapted to big data scenarios.  Goh \cite{goh03} proposed an approach using bloom filters which created a searchable index for each file containing trapdoors of all unique terms, but had the side effect of returning false positives due to the choice of data structure.  Curtmola \etal \cite{curtmola} worked off of this approach, keeping a single hash table index for all documents, getting rid of false positives introduced by bloom filters.  The hash table index for all documents contained entries where a trapdoor of a word which appeared in the document collection is mapped to a set of file identifiers for the documents in which it appeared.  Van Liesdonk \etal further expanded on this in \cite{liesdonk} with a more efficient search by using an array of bits where each bit is either 0 or its position represents one of the document identifiers.  These methods are generally faster, taking constant time to access related files, but are less provably secure, opening up new amounts of data to potential threat.  All of the mentioned methods only offer an exact-keyword search, leaving no room for user error through typos and cannot retrieve works related to terms in the query.

\subsection{Semantic Search}
Much of the work into searching semantically has been done in the context of searching the web \cite{survey, griddeddata, tonon}.  Some of these works, such as RQL by Karvounarakis \cite{rql}, require users to formulate queries using some formal language or form, which leads to very precise searching that is inappropriate for na\"{\i}ve or everyday users.  Others \cite{inquirus, semsearch} aim for a completely user-transparent solution where the user needs only to write a simple query with possible tags, while others still \cite{semanticsearch, shoe} aim for a hybrid approach in which the system may ask a user for clarification on the meaning of their query.  

All of these methods use some form of query modification coupled with an ontology structure for defining related terms to achieve their semantic nature.  In addition, these ontology structures often need to be large and custom-tailored to their specific use cases or domain, making them very domain-dependent and unadaptable to different areas.  Surprisingly, few of the works in this field offer a ranking of results, instead having the user choose from a potentially large pool of related documents.  

\subsection{Semantic Search over Encrypted Data}
Few works at the time of writing have combined the ideas of semantic searching and searchable encryption.  Works that attempt to provide a semantic search often only consider word similarity instead of true semantics.  

Li \etal proposed in \cite{li} a system which could handle minor user typos through a fuzzy keyword search.  Wang \etal \cite{wang} used a similar approach to find matches for similar keywords to the user's query by using edit distance as a similarity metric, allowing for words with similar structures and minor spelling differences to be matched.  Amini \etal presented in \cite{reseed,reseedjournal} a system for searching for regular expressions, though this still neglects true semantics for another form of similarity.  Moataz \etal \cite{moataz} used various stemming methods on terms in the index and query to provide more general matching.  Sun \etal \cite{sun} presented a system which used an indexing method over encrypted file metadata and data mining techniques to capture semantics of queries.  This approach, however, builds a semantic network only using the documents that are given to the set and only considers words that are likely to co-occur as semantically related, leaving out many possible synonyms or categorically related terms.  

\subsection{Clustering Methods for Searching}
The clustering hypothesis states that ``Closely associated documents tend to be relevant to the same requests'' \cite{clusterhypothesis}.  This idea has been expanded upon in many ways to form the body of research that investigates document clustering and its effects in information retrieval and searching.  Clustering has largely been used in two main ways: partitioning the central index into static shards, independent of user search queries, and clustering in a query specific manner based on the results of searches with the query \cite{LiuCroft}.  Solutions following the latter approach have the potential to outperform the static clustering approach \cite{Tombros}, they are largely impractical for large data sets.  

The former approach has been studied extensively, especially in the domain of web searching \cite{Yates, Google}, but these systems still demand a high computational cost to search over big data.  Relatively few works have specifically focused on the idea of clustering the central index into shards based on topics.  This idea was pioneered by Xu and Croft \cite{XuCroft}, who showed that making shards of a dataset's index more homogeneous (i.e. the contents of the shards are based around the same topic) improved the effectiveness of a system over standard distributed information retrieval. They used the k-means clustering algorithm with a KL-divergence distance metric to create the shards, then determine which shard should be searched by a query by estimating the likelihood that the query would come from the shard's language model.  Liu and Croft \cite{LiuCroft} expanded upon this by using more updated language modeling techniques to better smoothen their estimations.  However, neither of these works were appropriate for large scale data.

Kulkarni \etal \cite{Kulkarni} adapted these methods to larger scale datasets by performing the k-means clustering on a smaller sample of the dataset, then inferring from the documents' language models which shard those not included in the sample would belong to.  These works differ from ours in that they are only designed to operate on plaintext datasets.  Before this work, there was no attempt to create a topic-based clustering system that would operate on secured encrypted datasets.  Additionally, these models perform clustering on documents, whereas our work is designed to cluster terms from the documents, which was more effective given our encrypted approach.
\section{Architecture and Processes of S3BD} \label{sec:arch}
\system~has three primary architectural \underline{components}: the \textit{Client Application}; \textit{Cloud Processing Server}; and \textit{Cloud Storage}. Within those components, the system supports three major \underline{processes}, namely \textit{uploading documents}; \textit{clustering on the encrypted index}; and \textit{semantic search}. In this section, we first elaborate on the architectural components of \system, then explain the major processes.

\subsection{Overview of S3BD Architecture}
\begin{figure}
    \centering
    \includegraphics[width=0.65\textwidth]{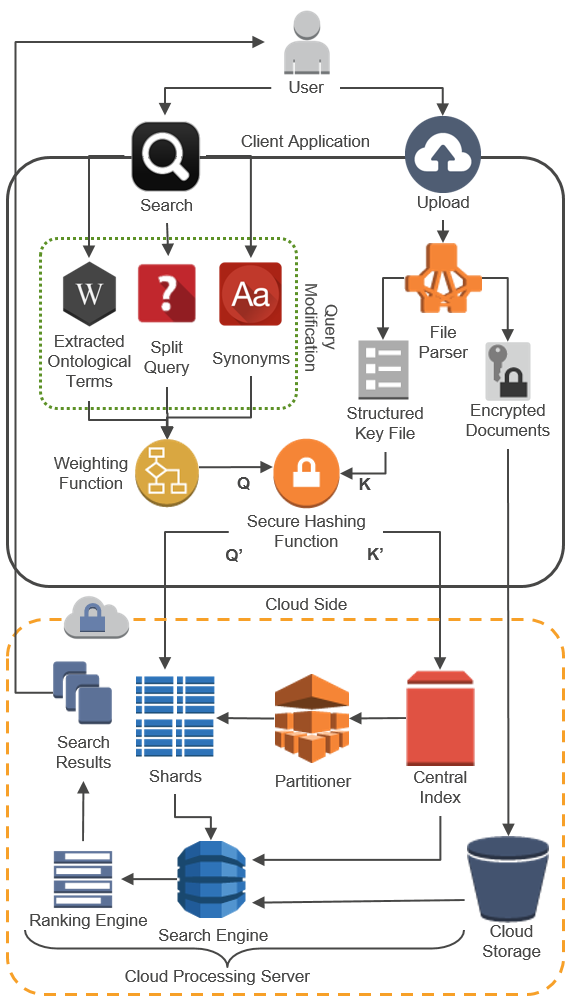}
    \caption{Overview of the \system~architecture and processes. Parts within the solid-line indicate components or processes at the user-end, deemed trusted. Parts in the dashed-line indicate those on the cloud-end, deemed untrusted.}
    \label{fig:architecture}
\end{figure}

Figure~\ref{fig:architecture} presents an overview of the components and processes in \system. In this figure, Client Application is a lightweight program hosted on the user's device and is the only component in the system deemed to be trusted. Cloud Processing Server and Cloud Storage are maintained by a third party cloud provider, thus, considered ``honest but curious''.  Our threat model assumes cloud components and the network channels are prone to external and internal attacks. 


The components in the architecture are described as follows:
\begin{itemize}

\item The \emph{Client Application} provides a user interface for uploading documents or to search over them in the cloud. It is also responsible for parsing and extracting information from plaintext documents and encrypting them before they are uploaded.

When the user requests to search, the system expands the search query with semantic data and transforms it to the secure query set (termed as a \emph{trapdoor}). The trapdoor is used for the search process on the cloud. The user then receives a ranked list of documents that can be downloaded and decrypted upon request. Client Application is also responsible for pre-processing queries to enable proactive searching on a subset of big data and achieve real-time search operation. 

\item The \emph{Cloud Processing Server} is responsible for constructing and updating the index and other related data structures during the upload process using encrypted data sent by the Client Application. 
Once the central encrypted index is built, it is clustered into shards to make search scalable for big data. As the clustering process is time consuming, it is performed in an offline manner as the dataset grows. 

Cloud Processing Server is also involved in the search process. It receives the user's search query and loads the relevant shards into memory. The shards are then searched to find and rank relevant documents. The highly-ranked documents are retrieved from the Cloud Storage and sent to the user.

\item The \emph{Cloud Storage} component is used to store the uploaded encrypted documents. Therefore, it does not see any representation of the user's query. Upon request by the Cloud Processing Server, the Cloud Storage can locate the documents and provide it to the user\footnote{Currently, our Cloud Storage relies on a single cloud. However, the architecture can potentially utilize multiple clouds for storage so long as the location of each document is provided.}. 

\end{itemize} 

Finally, it is important to note that each component exists \emph{per-user}. That is to say, while each Cloud Processing Server and Cloud Storage instance may exist on a single machine, components which hold data are assumed to exist separately for each user (e.g. each law enforcement agency). Thus, a separate central index and set of shards is held for each user. This is done to avoid bloating search times with operations to separate different users' files, and to avoid having users collude to attack other users.

\subsection{Upload and Parsing Process} \label{sec:upload}
Upon user request to upload a new document to cloud, the parsing process  extracts a subset of the terms and phrases in a document (called \emph{keywords}) to represent the semantics of that document. To preserve security, the keywords are encrypted along with the document before uploading to the cloud. The encrypted keywords are used to create (or update) the encrypted index structure on the cloud. 

A na\"{i}ve approach to extract keywords is to select all terms from a document excluding stopwords. However, previous works (\eg~\cite{s3c}) show that selecting few keywords that are semantically related to the document heavily reduces the storage overhead while still producing relevant search results. More specifically, the advantages of extracting a subset of keywords are three-fold. First, it maintains a low storage overhead for the central index. Second, assuming the central index is appropriately structured (\eg using hash tables~\cite{curtmola}), it makes the time complexity of updating the index with new documents nearly constant~\cite{curtmola}. Third, it increases the security of the central index by exposing fewer keywords to the external world. Thus, we use a key phrase extractor algorithm~\cite{Maui} to extract a number of keywords that represent the document's semantics. The composite (\ie multi-phrase) extracted keywords are split into individual distinct terms. This ensures that the central index contains encrypted versions of both the composite keyword and its components. 

Once keywords for a document are extracted, the frequency of their occurrences within the document is collected.  Then, the extracted keywords are deterministically encrypted~\cite{cryptdb}. Deterministic encryption is a method that always transforms a value (keyword) into the same encrypted token, similar to hashing.  In our implementation, we use the RSA deterministic encryption algorithm
\cite{RSA}
for this. Individual users who want to search the same dataset (e.g. law enforcement officers in a single agency) share RSA key pairs.

We use this method for the central index structure to allow for matching to encrypted query terms in the index, an integral part of the search process. It is worth noting that the frequency of keywords are maintained in plaintext in the central index. The use of homomorphic encryption
\cite{homomorphic:modern}
on the frequency data was considered, but the system needs to perform many operations on them and current implementations of fully homomorphic cryptosystems are too slow \cite{homomorphic:slow} to achieve our desired real-time response rate. 
Finally, the extracted keywords and their frequencies are integrated in a key file before uploaded to the cloud.

When the encrypted document and key file are received by the cloud server, the document is sent to the cloud storage block, while the key file information is added to the index. The cental index is stored as a mapping of encrypted terms to document IDs and the frequency at which they appeared in those documents.


\subsection{Topic-based Clustering Process} \label{sec:cluster}
\system~alleviates the search over the central encrypted index by clustering its terms into semantically related shards. In this research, we term this process as \emph{topic-based clustering}. The challenge is how to perform this type of clustering on the encrypted terms, because their meaning is lost due to encryption. One approach to overcome this challenge is to cluster terms based on their co-occurrences in documents, known as \emph{statistical semantics}. To achieve this, we adapt k-means clustering algorithm~\cite{kmeans} to cluster encrypted data at the keyword level.

K-means clustering algorithm allows us to cluster terms as long as the distance between two terms can be formulated. Distance between two keywords is defined as the semantic relatedness between them~\cite{XuCroft}. To adapt k-means for the encrypted data, we need to define the two main operations, namely picking initial means (also known as centroids) of clusters; and computing distance between the encrypted terms. These operations are discussed in the next subsections.

Once the clusters are built, they are used against search queries. To make the search scalable for big data, we define \emph{pruning} as to proactively search clusters that are semantically relevant to the query (\ie pruning irrelevant clusters). However, because the clusters are encrypted and the semantics are lost, the pruning cannot be achieved on the cloud.
For that purpose, in this section, we also develop a method, termed \emph{abstraction}, to sample the clusters into small abstracts that can be sent to and decrypted on the client-end. Upon search request, the abstracts are used on the client-end to prune the cluster and determine which clusters to be searched on the cloud.

\subsubsection{Initializing Centroids}

The first step to partition the central index into clusters with the K-Means clustering algorithm is to pick initial shard \emph{centers} (also known as centroids) to form the shards around them. A centroid is an entry in the central index which is essentially an encrypted keyword plus the list of documents it appears in. 

For effective search pruning, the shards should be distributed as evenly as possible, while maintaining semantic relationship. 
For that purpose, the centroids should represent diverse sections of the dataset. In fact, the initialization of centroids significantly impacts the resulting shards. 

A na\"{i}ve method for initializing centroids is to simply pick a number of centroids equal to the number of desired shards ($k$) randomly from the central index. This method can potentially result in keywords with very few associated documents being chosen as centroids. Because the co-occurance of keywords and centroids in documents determines the distance between them, the  na\"{i}ve method can potentially lead to formation of small shards (\ie shards with few elements). 

To avoid creating small shards (and uneven clustering), we propose a second method that ensures centroids have enough associated documents to attract other keywords. Our method for initializing centroids is to sort the keywords in the central index based on the number of their associated files, then choose the top $k$ terms as centroids. This ensures that keywords with low association are not chosen. However, this method can potentially lead to picking centroids with a high overlap of associated documents that can again cause uneven distribution of shards. Thus, ensuring centroids have a diverse set of associated documents is prioritized. 

The method we develop for choosing centroids operates on the sorted index and nominate keywords from the beginning that do not overlap, in their associated  documents, with previously nominated keywords. 
The algorithm to build centroids is mentioned in Algorithm~\ref{alg:centr}. The algorithm receives the number of clusters, denoted $k$, and the sorted central index structure as input parameters and determines the set of centroids and their associated documents, denoted $U$, as output.
\begin{algorithm}

\SetAlgoLined\DontPrintSemicolon
\SetKwInOut{Input}{Input}
\SetKwInOut{Output}{Output}
\SetKwProg {proc}{Procedure}{}{}
\SetKwFunction{main}{NominateCentroids}
\SetKwFunction{quant}{MeasureUniqueness}
\Input{$k$ and $central~index$ (with terms sorted by number of associated files)}
\Output{$centroids$ and $U$}
\proc{\main{k}} {
	$U \gets \emptyset$ \;
    $centroids \gets \emptyset$ \;
    \ForEach{$\omega \in central\; index$} {
    	$\Omega_\omega \gets \quant(\omega)$ \;
        \If {$\Omega_\omega \geq 1 $} {
        	// Nominate keyword to be a centroid \;
            $centroids.add(\omega)$ \;
            $U \gets U \cup I_\omega$ \;
        }
        \If {$centroids.count \geq k$} {
	    	\Return{centroids} \;
            \Return {U} \;
        }
    }
}

\proc{\quant{$\omega$}} {
	$unique \gets 0$ \;
    $duplicate \gets 0$ \;
    \ForEach{$documentID \in I_\omega$} {
    	\If {$documentID \in U$} {
        	$duplicate \gets duplicate + 1$ \;
        }
        \Else {
        	$unique \gets unique + 1$ \;
        }
    }
    \If {$duplicate > 0$}
    {$\Omega_\omega \gets unique \div duplicate$ \;}
    \Else {$\Omega_\omega \gets 0$ \;}
    \Return {$\Omega_\omega$}
}

\caption{Nominating Keywords as Centroids}
\label{alg:centr}
\end{algorithm}

For each keyword in the central index, we need to measure its uniqueness (Line 5 in Algorithm~\ref{alg:centr}). For that purpose, we develop a method to measure \emph{uniqueness} of a given keyword in the index structure (see Lines 16 to 33). Let $\omega$ a keyword and $I_\omega$ the set of documents associated with $\omega$ in the central index. Also, let $U$ the set of documents current centroids have appeared in. Then, we define \emph{uniqueness}, denoted $\Omega_\omega$, based on Equation~\ref{eq:uniq}. Uniqueness is also calculated in Line 28 of Algorithm~\ref{alg:centr}.

\begin{equation}\label{eq:uniq}
\Omega_\omega=\frac{|I_\omega \setminus U|}{|I_\omega \cap U|}
\end{equation}

In order to choose $\omega$ as a centroid, the number of documents unique to $I_\omega$ must be more than the number of documents in the intersection of $I_\omega$ and $U$ (\ie $\Omega_\omega \geq 1$, as indicated in Line 6). Upon choosing a keyword to be a centroid, the keyword and its associated documents are added to the set of current centroids (Lines 8 and 9 in Algorithm~\ref{alg:centr}). 



\subsubsection{Computing Distance between a Keyword and a Centroid} 

Once $k$ centroids are chosen, the distance from centroids to keywords in the central index needs to be calculated. With plaintext data, calculating distance is possible using techniques such as semantic graph~\cite{LiuCroft}. However, this technique is impossible when encrypted data are used. 

The clustering hypothesis states that keywords (also called terms) which co-occur (\eg in a document) can be considered related~\cite{clusterhypothesis}. This can be obtained even when terms are encrypted. Accordingly, we consider the co-occurrence of terms in a document as a reasonable metric for their similarity. That is, if two terms appear in the same document, they are considered related.

Recall that $I_T$ denotes the list of documents associated with term $T$ in the central index. Also, let $\theta(T, f)$ number of times (\ie frequency) term $T$ appears in document $f$. We define the \emph{contribution} of file $f$ to term $T$, denoted $\kappa(f,T)$, as the ratio of $\theta(T, f)$ to the total number of times term $T$ appears in the dataset (\ie frequency count of term $T$). Equation~\ref{contribution} shows the formal representation of contribution for file $f$.

\begin{equation} \label{contribution}
	\kappa(f,T) = \frac{\theta(T, f)}{\sum\limits_{j \in I_T}{\theta(T, j)}}
\end{equation}

Let $\gamma_i$ be centroid of cluster $i$. We define \emph{contribution of term $T$ and file $f$ to cluster} $i$ (denoted $K(T,f,\gamma_i)$) as the ratio of sum of the frequency of $T$ in file $f$ and in $\gamma_i$ to the sum of frequency count of $T$ and $\gamma_i$. Equation~\ref{eq:contrbetween} shows the formal definition of $K(T,f,\gamma_i)$.

\begin{equation} \label{eq:contrbetween}
	K(T,f,\gamma_i) = \frac{ \theta(T,f) + \theta(\gamma_i,f)}{ \sum\limits_{j \in I_T}{\theta(T, j)} + \sum\limits_{p \in I_{\gamma_i}}{\theta(\gamma_i, p)} } 
\end{equation}

Then, we define \emph{cooccurence} of term $T$ and $\gamma_i$ through file $f$, denoted $\rho (T,\gamma_i,f)$, as the ratio of $\kappa(f,T)$ to $K(T,f,\gamma_i)$. Equation~\ref{eq:coccur} shows the formal definition of cooccurence. 

\begin{equation}\label{eq:coccur}
	\rho (T,\gamma_i,f)=\frac{ \kappa (f,T) }{K(T,f,\gamma_i)}
\end{equation}

To represent the similarity between term $T$ and centroid $\gamma_i$, we compute the distance, denoted $d(\gamma_i, T)$, based on Equation~\ref{distance}. In this equation, we iterate through the list of documents that are associated with the term. For each document, we consider the contribution of that document to term $T$ and the cooccurence of $T$ and $\gamma_i$ through $f$. We use logarithm to limit the impact of the cooccurence factor.
\begin{equation} \label{distance}
	d(\gamma_i, T) = \sum_{f\in I_T} \kappa (f,T) \cdotp \log_{10}{(\rho(T,\gamma_i,f))}
\end{equation}



\subsubsection{Evening Shards Sizes}
Once the similarity between each centroid and each term is computed, terms can be distributed to their proper shards.  An initial approach for distribution is to assign each term to the shard with the centroid that the term has the maximum similarity with. This approach, however, can potentially lead to uneven shard distribution and subsequently inefficient search pruning.

To avoid uneven clustering, we limit the growth of each shard so that it can only hold up to a certain amount of its closest terms. Ideally, all clusters should end with an equal size of $\frac{|I|}{k}$ in which $|I|$ is the total number of terms in the central index. Accordingly, we constrain the growth of each shard to $\frac{\alpha\cdotp |I|}{k}$. Parameter $\alpha$ is determined to be greater than 1 (\ie $\alpha > 1$) to cover the dynamism of a natural language but does not allow a shard (\ie topic) to dominate the clustering. In our implementation, we considered $\alpha =2$. 

In a circumstance that a shard reaches its threshold, we disassociate the furthest term from the shard's centroid. Then, we assign it to the closest shard that has not yet reached to its threshold.

Once the shards are initialized with the terms, based on the afore-mentioned approach, we iteratively reorganize them to produce shards that are more centered around a topic. We define the \emph{average term} of shard $\pi$ as the closest term to the average distance of all terms from the current centroid of shard $\pi$. In each iteration, the clustering process of \system~calculates the average term in each shard, chooses it as the new centroid, and forms a new shard around it.  

Ideally, the iteration would continue until the shards' composition stabilizes. That is, when there is no alteration of terms during an iteration. However, in practice, as the iterative clustering is a computationally expensive operation, we limit the iterations until shards are minimally altered. In our implementation, we realized that we generally reach to the stable state in five iterations.

\subsubsection{Shard Abstraction}
Because the shards on the cloud processing server are all encrypted, it is impractical to identify shards related to search query and perform pruning on the cloud. Thus, we need a method to identify appropriate shards to search over, for a given user query. We propose abstracting the shards into tiny unencrypted samples that are processed on the client premises. 
These abstracts are used against search queries to navigate search to only shards contain relevant search results.


As a centroid shows centrality of a shard, it is more indicative of the shard's general topic, hence, can be used to form the abstract. However, each centroid is only a list of documents and cannot directly be used in abstracts. Therefore, the system chooses terms from the documents associated with a centroid to build that shard's abstract.  In particular, it chooses the most frequent term from each associated document of a centroid.  

Each abstract, which is a small set of encrypted terms, is sent to the client machine. The abstracts are decrypted on the client machine and compared to the search query terms through a semantic similarity metric \cite{wupalmer}. Although abstracts are small, using semantic similarity of search query to abstract terms enable identifying most relevant shards. Then, the search query is only compared against those identified shards in the cloud.
\subsection{Search Process} \label{sec:search}
The search process consists of three main phases: \emph{abstract comparison}; \emph{query modification}; and \emph{searching and ranking}.  The Client Application is responsible for the first two phases, while the third happens on the Cloud Processing Server.

In summary, \system~first performs pruning by comparing the query against the abstracts to determine the shards that need to be searched. This information is then sent to the Cloud Processing Server to load the appropriate shards into the memory as soon as possible. Meanwhile, the client application semantically expands and the search query,  encrypts it, and sends it to the cloud processing server. 

Once the Cloud Processing Server has all of the necessary information, it finds and then ranks relevant documents from the shards specified in the abstract comparison phase. The result list is then sent back to the user. The document selected by the user is downloaded and decrypted on the client machine. 

\subsubsection{Comparing Queries Against Abstracts}
As the terms in the abstracts are semantically linked to the topics in corresponding shards, comparing the query to the abstracts let \system~detect which shards are appropriate to search.  This comparison is carried out, in the first phase of search, by obtaining semantic distance of the terms in the query to the terms in the abstracts using the WuPalmer word similarity metric~\cite{wupalmer}. For this purpose, WuPalmer computes the semantic similarity between two words by evaluating the distance from one word to the other in a large semantic graph, returning a normalized value between 0 and 1.

Using WuPalmer to compare the query against all abstracts allows the system to rank the abstracts based on relevance to the query. As the abstracts represent the shards, the ranking can identify relevant shards to be searched. 

The number of shards chosen determines the trade-off between search time overhead and search comprehensiveness that can be decided based the user discretion. In fact, choosing a higher number of shards consumes more memory and increases the search time, conversely, a low number of shards ignores searching less relevant parts of the dataset that can include desirable results. 

Once a sufficient number of shards have been chosen, the cloud processing server is notified and begins loading those shards into memory. The loading time is overlapped with the Query Modification step (explained in the next section) at the client end. 

\subsubsection{Query Modification}
Query modification is meant to inject semantic information into the query. This phase starts with the user entering a plaintext query, denoted as $q$, into the Client Application, after which it goes through three steps: \emph{query parsing, semantic expansion,} and \emph{weighting}. These steps result in forming a \emph{query set}, hereafter noted as $Q$. 

The goal of query parsing is to refine the search query and split it into smaller tokens or sub-phrases. 
To refine $q$, we first remove all stop words~\cite{stopwords} (\eg articles and prepositions).
If the query is multi-phrase, we then split it into parts. 
The reason for splitting is twofold. \emph{First,} because some documents may partially match with the query. \emph{Second}, because portions of a query cannot be derived from the encrypted query. Hence, we split $q$ and create all tokens and sub-phrases of it before encryption.
Once this step is complete, $Q$ consists of $q$, its split parts, and its sub-phrases.

The goal of semantic expansion is to add terms related to the query into the query set, thus, enabling \system~to search for semantically related results. In order to achieve this, \system~injects semantic data extracted from an ontological network~\cite{wikipediabased}. 

One approach to extract semantic data is to perform a synonym lookup (\eg through an online thesaurus) for each member of $Q$ (termed $Q_i$) and add the results to $Q$. 
However, this approach alone does not produce concepts that are semantically related to the user's query, but are not synonymous.  To cope with this problem, \system~needs to pull related terms from conceptual ontological networks~\cite{wikipediabased}.  

For the development of \system, the elements of $Q$ are used to pull entries from Wikipedia, as an instance of an onthological network. Keyphrase extraction is performed on the entries to get conceptually related terms and phrases, which are added to $Q$.  

In the search results, documents that include phrases exactly matching query terms are deemed more relevant. For that purpose, in the weighting step, we assign weights, ranging from 0 to 1, to the elements of $Q$ as follows:

\begin{itemize}
\item As the documents that include the whole originial query have top priority, we assign the maximum weight of $1$ to $q$.

\item Documents that include parts of the query (\ie $Q_i$) are more relevant than those including terms derived from the query. As such, we assign $1/n$ weight to results of the query parsing where $n$ is the number of parts in the search query.

\item Documents including related terms derived from the query have the lowest relevance. Accordingly, the related terms obtained for each  $Q_i$ should be assigned the lowest weight. Let $W(Q_i)$ be the weight of $Q_i$. Then, terms derived from $Q_i$ are weighted as $W(Q_i)/m$ where $m$ is the number of derived terms from $Q_i$.
\end{itemize}

\subsubsection{Searching and Ranking}
Once the query set $Q$ is built, its elements are deterministically encrypted~\cite{cryptdb} to create the trapdoor $Q'$. The trapdoor is then sent to cloud processing server to perform the search and ranking of the result set. On the cloud processing server, each element of $Q'$ is checked against a union of the loaded shards, denoted $\Pi$, to compile a collection of documents, denoted $C$, that are potentially related to the query. 
Once $C$ is compiled, the documents are ranked and then sent to the user.

The search operation has to be agnostic about semantics of the query and the dataset. Okapi BM25~\cite{okapi} is a search and ranking algorithm for unencrypted data that functions in the  meta-data level and provides the required data agnosticism. Okapi BM25 operates based on the list of keywords provided to it. However, it cannot differentiate between elements of the query set ($Q'$). We extend the idea of Okapi BM25 to include encrypted data and to consider the weighting of elements in $Q'$. 

Based on Okapi BM25, a document's rank for a given query is considered to be a function of the following three factors:
\renewcommand{\theenumi}{\Alph{enumi}}
\begin{enumerate}
\item \emph{frequency} of query term $Q_i$ in document $d_i$, denoted $f(Q_i, d_i)$. In \system, $f(Q_i, d_i)$ can be obtained by looking up the encrypted query element $Q'_i$ in $\Pi$.

\item \emph{Inverse Document Frequency} (IDF) of query term $Q_i$ across collection of documents ($C$). Let $N$ be the total number of documents in $C$, and $n(Q_i)$ be the total number of documents containing $Q_i$.  Then, Equation~\ref{IDF} defines the IDF for $Q_i$. 
\begin{equation} \label{IDF}
IDF(Q_i) = \log \frac{N-n(Q_i) + 0.5}{n(Q_i) + 0.5}
\end{equation}

In $\Pi$, we keep the frequency of each term in each document. Therefore, $n(Q_i)$ can be obtained by summing up all the frequencies of $Q'_i$ in $\Pi$. 

\item \emph{Document Length Normalization} (DLN) that removes the effect disparity in documents' length. Let $\delta$ be the average length of all documents in $C$, and $\beta$ be a parameter that determines the impact of the DLN factor. Equation~\ref{dln} formally defines DLN for $d_i$. In this work, we considered $\beta=0.75$. 
\begin{equation} \label{dln}
DLN(d_i) = (1 - \beta + \beta \cdotp \frac{|d_i|}{\delta})
\end{equation}

In \system, we maintain the length of uploaded documents. As such, we can obtain $\delta$ and $|d_i|$ to calculate DLN for $d_i$.

\end{enumerate}

A rank is defined as the sum of scores given by each $Q_i$. To consider the weighting scheme of $Q$ in ranking, each score is adjusted by considering the weight of $Q_i$ (denoted $W(Q_i)$).
Equation~\ref{eq:okapi} shows the formal representation of rank of document $d_i$ for query set $Q$ (denoted $r(d_i, Q)$).
In this equation, $alpha$ is a parameter that determines the impact of the frequency factor. Our initial experiments show that $\alpha=1.2$ provides an accurate ranking, thus, we use this value in our implementation.
\begin{equation} \label{eq:okapi}
r(d_i, Q) = \sum_{i=1}^{n} IDF(Q_i) \cdotp \frac{f(Q_i, d_i) \cdotp (\alpha + 1)}{f(Q_i, d_i) + \alpha \cdotp DLN(d_i)} \cdotp W(Q_i)
\end{equation}

The cloud processing server computes Equation~\ref{eq:okapi} for all encrypted documents in the collection $C$ against $Q'$. To exploit parallelism implicit in the cloud system, $C$ can potentially be compiled using a mapreduce approach, mapping each shard to a separate process, spreading $\Pi$ across multiple machines. Because individual documents can be represented across multiple shards, an additional process would need to combine the scores accumulated from different processes for each document. Once $C$ is compiled, its members are ranked in descending order, and a list of document identifiers are sent to the client to be picked.


\section{Security Analysis} \label{sec:security}

\system~provides a trustworthy architecture for storing confidential information securely in clouds while maintaining the ability to search over them.  Our threat model can be defined as follows. Our system architecture is divided into three major components that live either in the cloud or on the user's machine (as seen in Figure \ref{fig:architecture}). Only components and processes that exist on the user's machine (i.e. the Client Application) are considered to be trusted, meaning we can store and access plaintext data there. Keeping the user's machine trusted is a reasonable assumption in the real world, as it can be kept with minimal exposure to outside attackers.

Components in the cloud are considered untrusted and susceptible to adversaries. We consider these attackers to be either external (i.e. an unaffiliated party who wishes to learn about the dataset) or internal (i.e. a party with access to the cloud who wishes to see the unencrypted dataset). Our threat model assumes that these adversaries may intend to attack the communication streams between the Client Application and Cloud Processing Server, and between Cloud Processing Server and Cloud Storage, as well as the cloud machines themselves. To explain exactly what threats the attackers pose to the encrypted data, we introduce the following definitions:



\textit{History}: For a multi-phrase query $q$ on a collection of documents $C$, a history $H_q$ is defined as the tuple ($C, q$).  In other words, this is a history of searches and interactions between client and cloud server.

\textit{View}: The view is whatever the cloud can actually see during any given interaction between client and server.  For our system, this includes the encrypted index and all shards $I$ over the collection $C$, the trapdoor of the search query terms (including its semantic expansion) $Q'$, the number and length of the files, and the collection of encrypted documents $C'$.  Let $V(H_q)$ be this view.

\textit{Trace}: The trace is the precise information leaked about $H_q$.  For \system, this includes file identifiers associated with the search results of the trapdoor $Q'$ and unencrypted weight information from $Q'$.  It is our goal to allow the attacker to infer as little information about $H_q$ as possible.

Because our threat model assumes a secure user machine, the View and Trace encompass all that the attacker would be able to see. For encryption on the plaintext documents being searched, we use a probabilistic encryption model, considered to be the most secure form of encryption \cite{cryptdb}. Hence we can infer that, because probabilistic encryption does not use a one-to-one mapping, $C'$ is not susceptible to dictionary-based attacks\cite{dictionaryAttacks}, and secure so long as the attacker can not access the keys (stored only on the user's machine).

$I$, in the View, only shows a mapping of a single deterministically encrypted term or phrase to a set of file identifiers with frequencies, meaning a distribution of encrypted terms to files could be compiled, but minimal data could be gained from the construction.  Similarly, $Q'$ only shows a listing of encrypted search terms with weights.  

The addition of the weights to $Q'$ could potentially enable the attacker to infer which terms in the trapdoor were part of the original query. Even in this case, the attacker can at most get the deterministically encrypted query.  Additionally, the expansion of the query to include semantic data adds noise that can mislead attackers from the original user query.

However, we must consider the small possibility that, if the attacker is able to obtain the keys used for deterministic encryption from the user's side, they could in theory build a dictionary of all words in the vocabulary $V$ that the documents are comprised of, mapped to their encrypted counterparts, and reconstruct $I$ in plaintext.  In this scenario, the attacker could put together the terms that the documents are comprised of, but since $I$ carries no sense of term order, they could not reconstruct the entire file.  Additionally, only a small portion of important terms and phrases from each document are given, meaning the attacker would only be able to ascertain how many times those specific terms and phrases were in the document.

An attacker monitoring the process during a search could see the resultant file identifiers that are associated with the given $Q'$.  This would show an encrypted history as ($C', Q'$).  However, since the attacker would not be able to discern the query (without the use of the above dictionary), this data would be of little use.

Attackers could also potentially attempt to alter data in $C'$.  These attacks, however, could be recognized as the Client Application would not be able to decrypt them.
\section{Evaluation} \label{sec:eval}

\subsection{Experimental Setup}
We have implemented a prototype of \system~which is available to the general public\footnote{A binary of \system~can be downloaded from \url{http://hpcclab.org/products/S3BDJars.zip}}. The implementation has been the platform for all experiments in this research. All experiments were carried out on Amazon EC2 cloud Virtual Machines (VMs). We ran both client application and cloud processing server on separate Amazon EC2 VM instances. In particular, we used m4.xlarge VM instance type to host client application and i2.xlarge VM instance type to host cloud processing server.

We evaluate two major aspects of \system, namely its \emph{performance} and its \emph{accuracy}. Performance specifically refers to the amount of time it takes to perform a search, while accuracy refers to the relevance of its search results.

To evaluate the performance of \system~with big data, we tested it utilizing portions of the Common Crawl Corpus dataset~\cite{comcrawl} from Amazon Web Services. The dataset consists of approximately 151 terabytes of text data obtained from an extensive web crawl. We parsed files within the dataset to create sample subsets (termed samples) of varying sizes. Each file in the dataset includes text from multiple web pages, hence, we split those files to create a document for each web page. To generate each sample, we randomly choose files until we reach the desired size for the sample.

We evaluate performance of \system~by analyzing the time to search over varying numbers of shards in the cloud. The results of this evaluation helps us determining the appropriate number of shards to create. 

Once we determine the appropriate number of shards, we compare the search time of \system~against that of our previous work (S3C) \cite{s3c}, as a baseline. S3C performs a similar style of semantic search without use of clustering and pruning. To assure that the performance results for \system~is not affected by the temporal performance variations of cloud VMs, we run each query 20 times and report the mean and 95\% confidence interval of the results. 

To evaluate the relevance of \system, we evaluated it using the Request For Comments (RFC) dataset~\cite{rfcdataset}. RFC is a collection of 6298 documents (247 MB) regarding notes on Internet development topics. The reason we choose RFC dataset is that it is domain-specific and small enough to manually verify accuracy of search results. We compare the results of \system~to the baseline (S3C) and to a version of S3C that operates on non-encrypted data~\cite{s3c}. 

In addition to these two major aspects, we analyze the storage overhead incurred by storing the central index on the Cloud Processing Server.  To that end we show the size of the central index as the dataset increases in size.

It is noteworthy that, as we use two different datasets for our evaluations, we generate two different sets of benchmark queries based on the nature of the datasets. The benchmarks are explained in the respective subsection for each experiment.

\subsection{Evaluating Performance of \system}
\subsubsection{Benchmark Queries to Evaluate Performance}
Perfrormance evaluations are carried out based on 10 benchmark search queries, shown in Figure~\ref{fig:perfbenchmarks}. 
Queries were chosen after manual analysis of the samples and determining topics of their documents. It is noteworthy that the wording of the benchmark queries does not impact the performance of the system.


\boxfig{
\texttt{Government News Report} \\
\texttt{Social Media Feed} \\
\texttt{Encryption in Cloud Server} \\
\texttt{City Travel Blog} \\
\texttt{Online Shopping Network} \\
\texttt{Free Recipe Index} \\
\texttt{Online Weather Service} \\
\texttt{Celebrity Fashion Catalog} \\
\texttt{Academic Conference Papers} \\
\texttt{Sports Broadcasting Network}
  \vspace{0.03in}
\caption{Benchmark queries used for evaluating search performance on the Common Crawl Corpus dataset.}\label{fig:perfbenchmarks}
}

\begin{figure}
  \centering
  \includegraphics[width=0.7\textwidth]{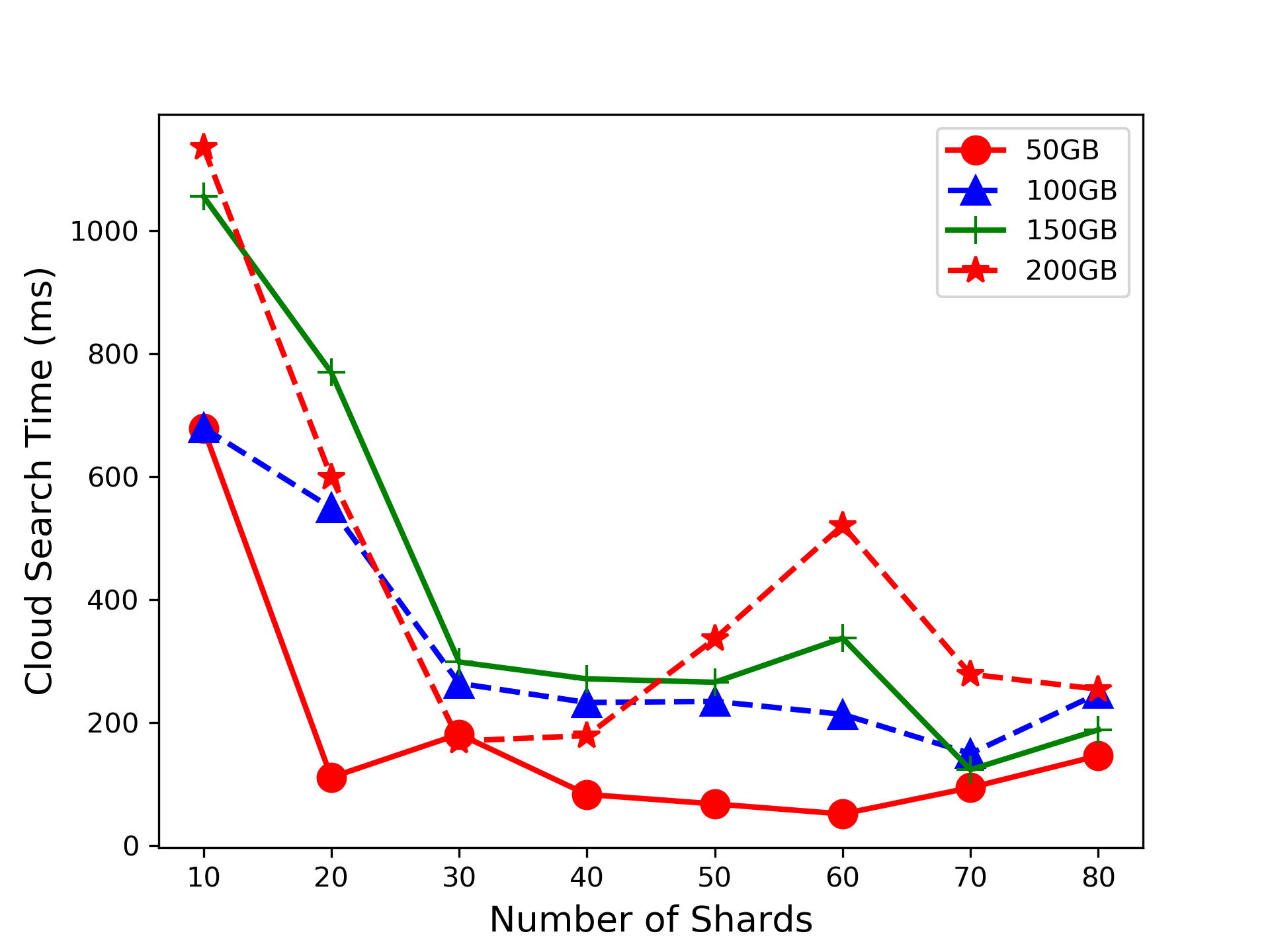}
  \caption{Time taken to search over the union of chosen shards ($\Pi$) on the cloud. The horizontal axis shows the number of shards the dataset sample is partitioned into and the vertical axis shows the average search time (in milliseconds). The search time includes time to find documents in $\Pi$ that match the trapdoor ($Q'$) and rank them. Each data point is the  average of searching each benchmark query 20 times.}
  \label{fig:shards}
\end{figure}

\subsubsection{Finding the Appropriate Number of Shards}
The challenge in \system~is how to determine the number of shards that should be created to provide the best trade-off between search performance and search accuracy. In fact, creating a lower number of shards potentially implies higher search accuracy, because each shard covers a larger subset of the dataset. However, searching a larger portion of the dataset lowers the performance, because less of the dataset is pruned.
On the other hand, creating a high number of shards does the inverse. That is, it improves search performance but potentially lowers accuracy. 

To handle the trade-off between the search performance and the search accuracy, in this experiment, we utilize the idea of Pareto front analysis~\cite{18xiangbo} to understand the relation between these factors and find the number of shards that satisfies both objectives at the same time. For that purpose, we compare the performance, in terms of cloud search time, across different numbers of shards, representing accuracy. 

The result of the experiment is shown in Figure~\ref{fig:shards}. The vertical axis, in this figure shows the time taken to search on the cloud and the horizontal axis shows the various numbers of shards created. To assure that our analysis is comprehensive and is not bound to a certain dataset size, we conduct the experiment with samples of different sizes --- from 50 GB to 200 GB. 

As we can see in Figure~\ref{fig:shards}, the time to search decreases as more shards are formed. Important to note is that declines in search time cease being substantial for 100, 150, and 200 GB past 30 shards. On the other hand, creating fewer shards yields remarkably high search times for larger datasets. As this pattern is consistently observed for samples of different sizes, we can conclude that partitioning datasets into  30 shards provides an ideal trade-off between search performance and accuracy of \system.

\subsubsection{Shard Distribution and Variance}

\begin{figure}
  \centering
  \includegraphics[width=0.7\textwidth]{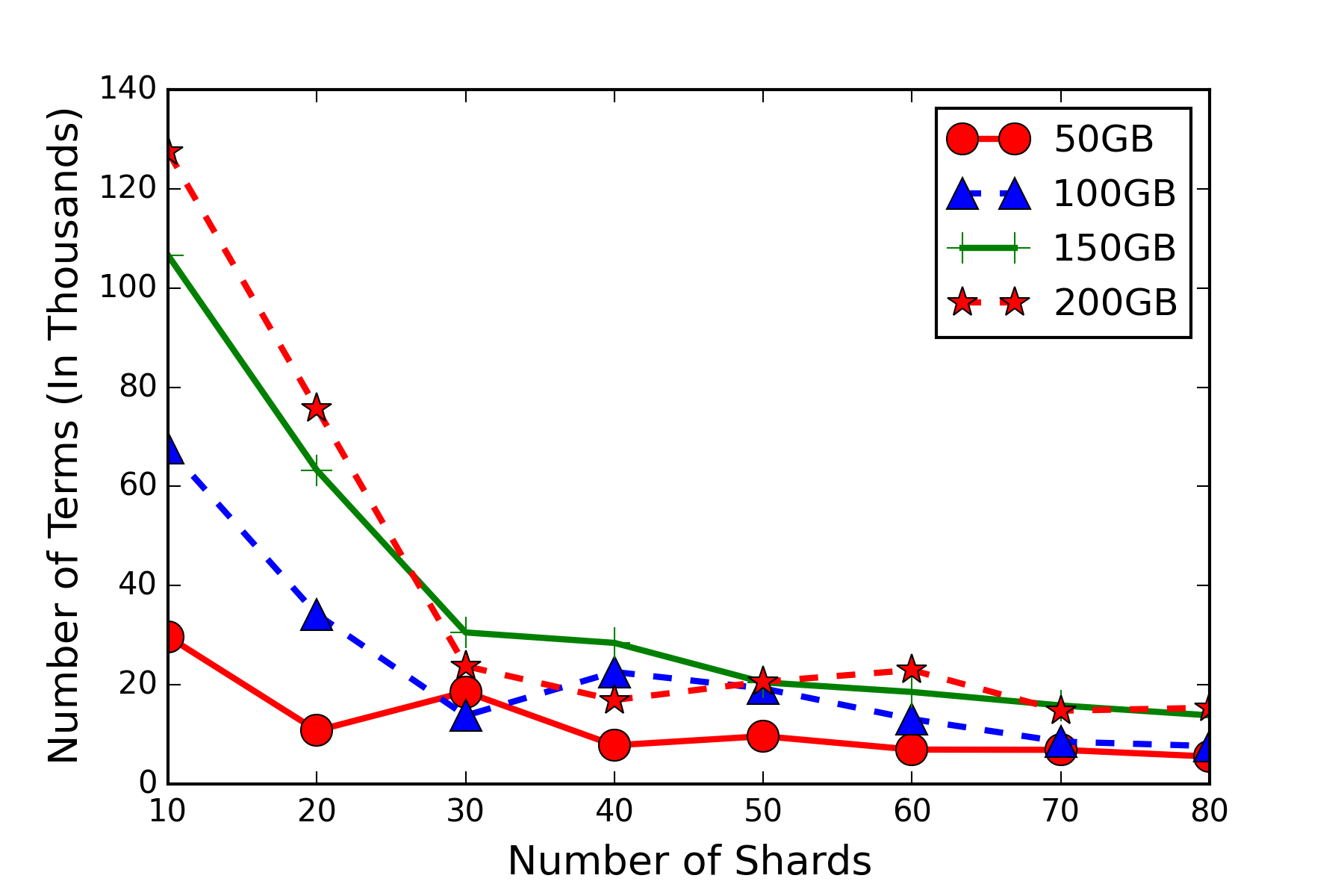}
  \caption{The average number of terms (\ie size of) the searched shards ($\Pi$). The horizontal axis shows the number of shards the dataset sample is partitioned into. The vertical axis shows the average number of terms of the searched shards.}
  \label{fig:avgterms}
\end{figure}

\begin{figure}
  \centering
  \includegraphics[width=0.7\textwidth]{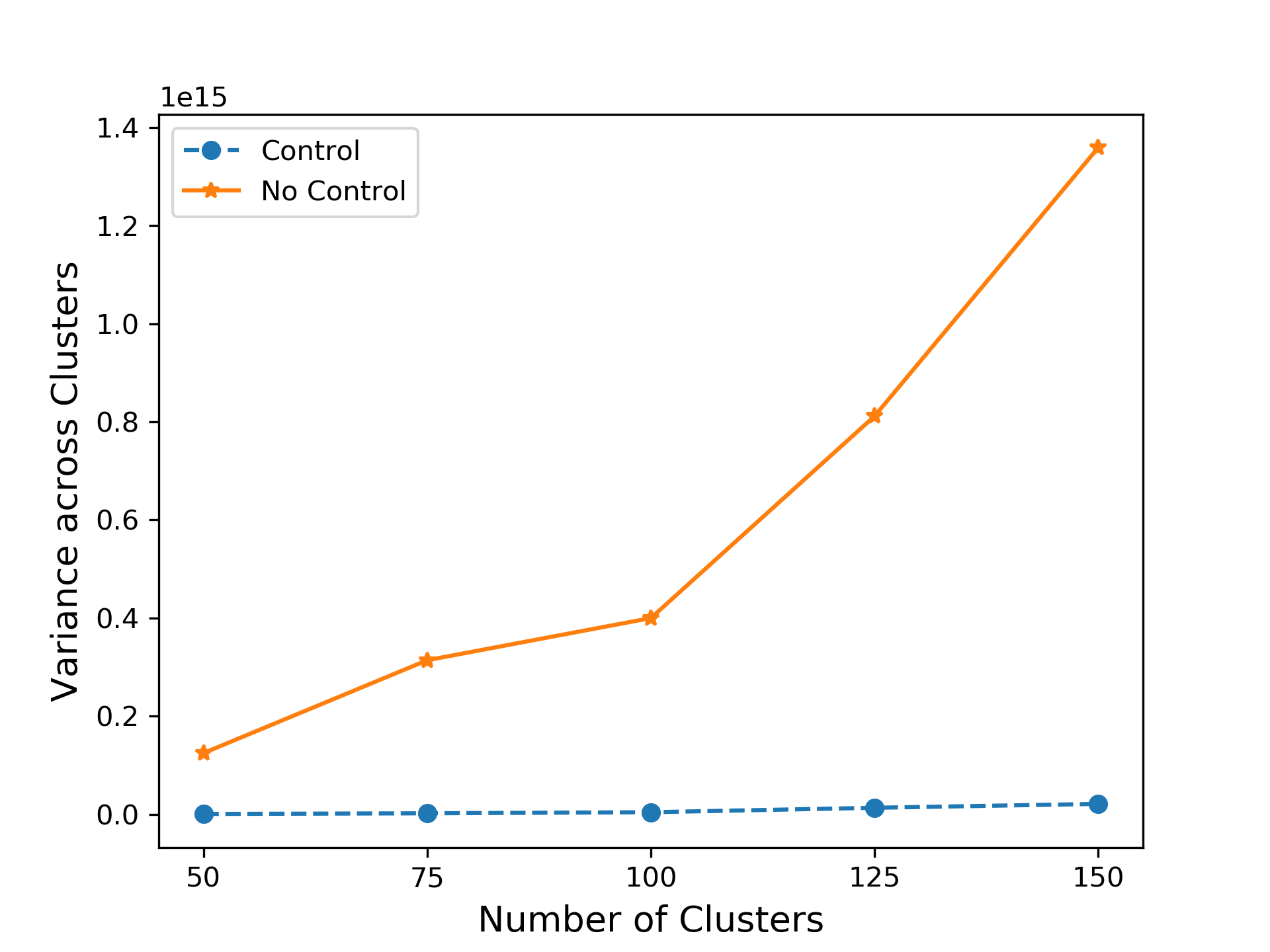}
  \caption{The variance in cluster sizes for controlled and non-controlled clustering. The vertical axis shows the variance, with cluster size measured in the number of documents represented in the cluster. The horizontal axis shows the size of the dataset.}
  \label{fig:clustersizevar}
\end{figure}

Interestingly, the search time does not strictly decrease as more shards are formed.  This can be seen primarily in the spike in search time for the 150 and 200 GB samples with 60 shards. Our analysis shows that this is attributed to uneven shard distribution. In fact, the size of a shard plays an important role in determining its search time. For a given query, the system can potentially determine to search among the smallest shards, when few shards are created. Inversely, for the same query with more shards created, it is possible to search among the largest shards, despite the fact that the average shard size is smaller.

We thus determine that maintaining consistent cluster sizes (low variance) is important for maintaining consistent search performance and accuracy. With \system, we introduce a method for controlling cluster size variance through diversifying cluster centroids and enforcing a maximum cluster size. To measure the effects of this, we analyze the variance in cluster sizes produced by the clustering algorithm with and without this cluster control. Results are seen in Figure \ref{fig:clustersizevar}. The Vertical axis of the figure shows the variance in the sizes of the shards, while the horizontal axis shows the number of shards created.

In the figure, we observe that the variance in the non-controlled clusters is substantially greater than that in the controlled clusters. We can thus infer that, while our control measures leave some variance in cluster sizes, it will lead to more consistent search performance and accuracy.

To further show the relation between search time and the size of searched shards (\ie number of terms in the shard), we analyzed the number of terms in the shards chosen to be searched, during performance evaluations, in Figure \ref{fig:avgterms}. Vertical axis of the figure shows the average number of terms in a searched shard, and horizontal axis shows the number of shards created. The rest of setup for this experiment is the same as those for Figure~\ref{fig:shards}.

By comparing the two figures, we observe that the number of terms in shards is correlated with the respective search times; both follow similar patterns. In addition, Figure~\ref{fig:avgterms} shows that creating more shards does not necessarily impact the average number of terms in searched shards nor does it improve search times. These justifications support our conclusion that 30 shards is an appropriate number of shards to be created in \system. 


\subsubsection{Performance Comparison of \system~Versus S3C}

\begin{figure}
  \centering
  \includegraphics[width=0.7\textwidth]{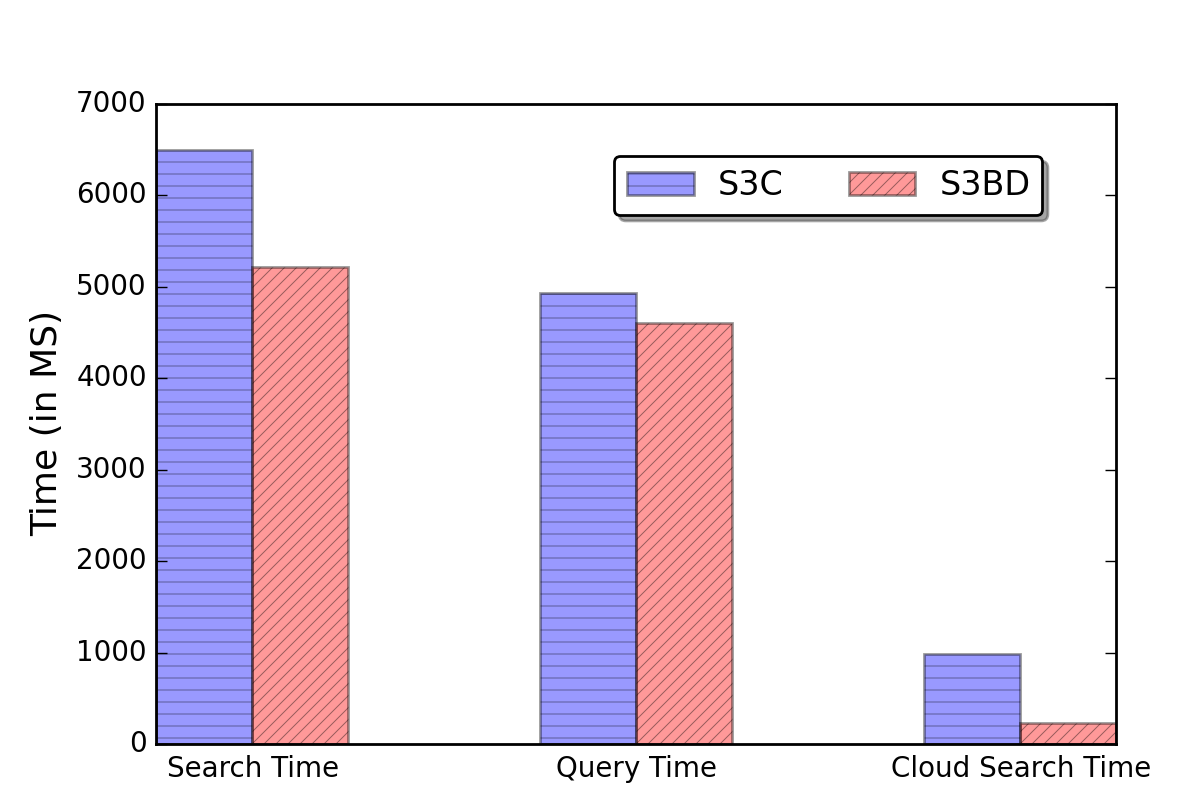}
  \caption{Detailed comparison of the performance of search components in \system~versus S3C. In this experiment, the dataset sample is 50 GB, clustered into 30 shards in \system. Query Time refers to time for query modification; Cloud Search Time is the times to perform searching and ranking; Search Time is the collective time to search. The vertical axis shows the amount of time taken (in milliseconds) for the corresponding component.}
  \label{fig:s3c_v_s3bd}
\end{figure}

\begin{figure}
  \centering
  \includegraphics[width=0.7\textwidth]{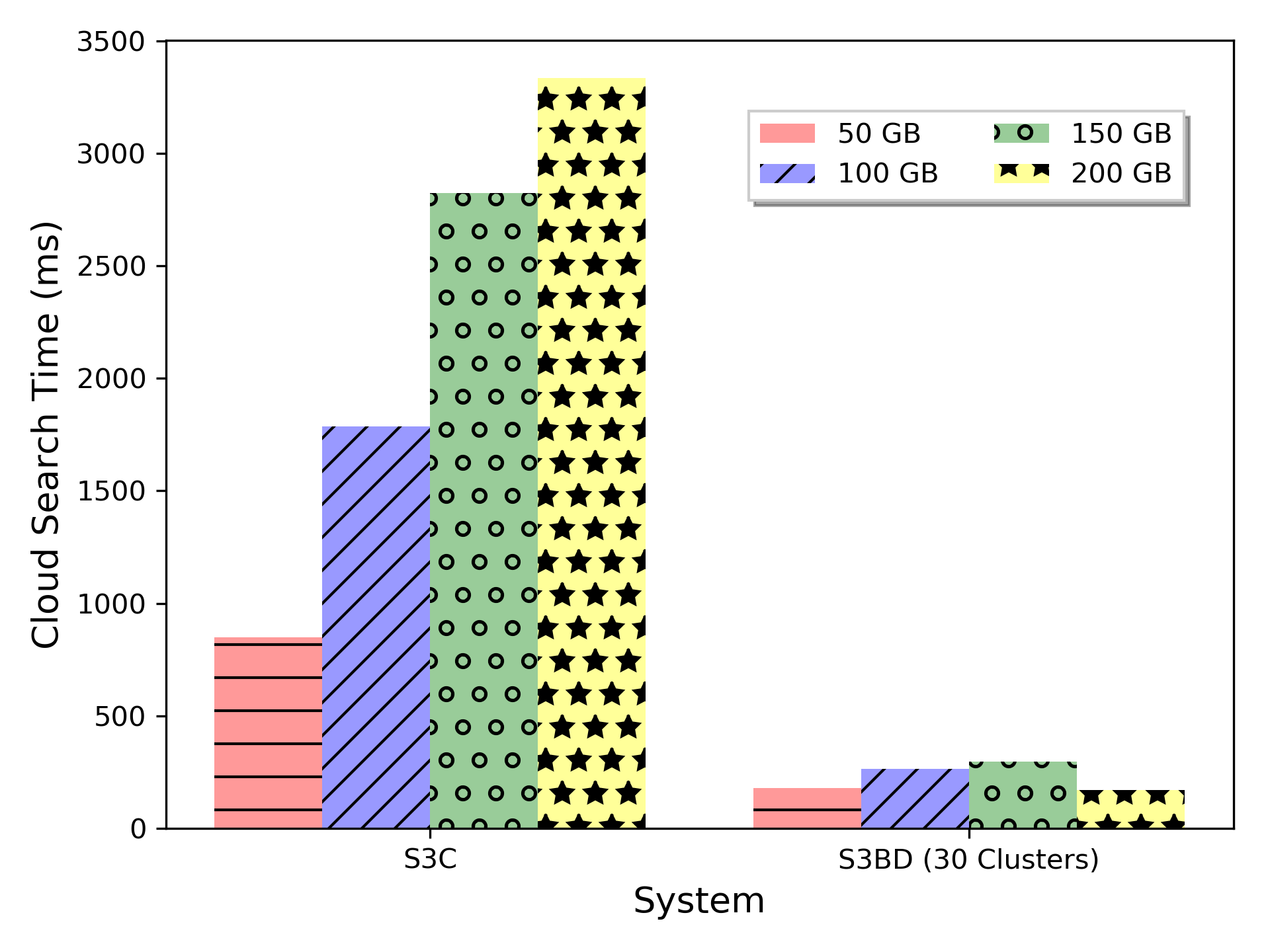}
  \caption{Comparison of the performance of S3C and S3BD across different dataset sizes. Each bar represents a different dataset size, and the vertical axis represents the time taken (in milliseconds) to perform searching and ranking on the cloud.}
  \label{fig:1_shard_vs_30}
\end{figure}

To show improvements in search time, we compare the search times of \system~against S3C, the earlier work in the literature. In this experiment, we first compare the components of the overall search time (query modification time and cloud search time) between the two systems using a dataset size of 50 GB. We then compare the cloud search time for both systems across various dataset sizes.  In accordance with the conclusion of previous experiment, we configured \system~ to cluster the sample into 30 shards. 

Figure \ref{fig:s3c_v_s3bd} shows the overall search time as well as the time of the two major actions involved in the search, namely query modification and searching and ranking on the cloud.  
According to the figure, the total search time of \system~is 20\% less than S3C. While query modification time is not significantly different, the figure expresses that the difference is due to cloud search time, which is 77\% less for \system.

Figure \ref{fig:1_shard_vs_30} shows time taken to search and rank on the cloud for the different systems using increasing dataset sizes. As shown by the figure, the search time increases at a higher rate for S3C than \system. Additionally, all search times for \system~are substantially lower than S3C, with time for 200 GB of data at $\sim$95\% less for \system. Because of this, \system~shows more promise for scalability to larger dataset sizes.


\subsection{Evaluating Overhead of \system}
\begin{figure}
  \centering
  \includegraphics[width=0.7\textwidth]{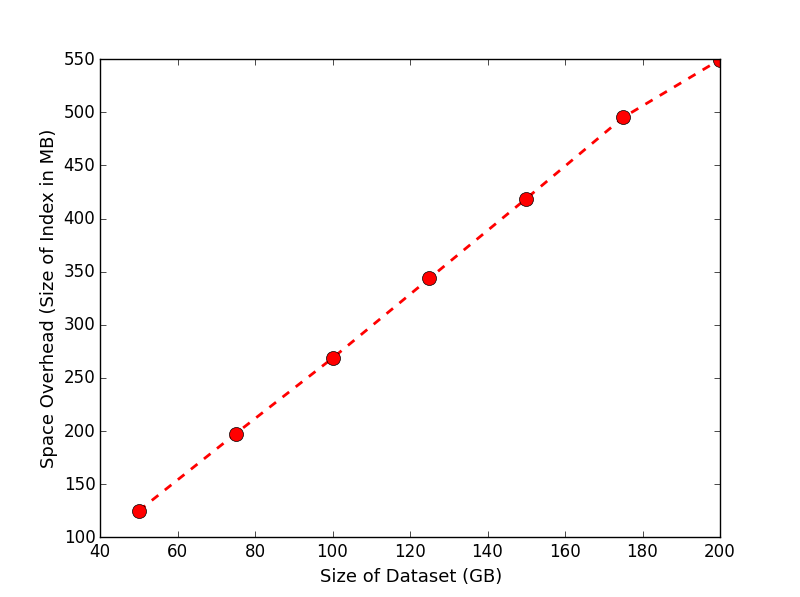}
  \caption{The size of the central index as the dataset increases in size.  The horizontal axis plots the size of the dataset used in
gigabytes, while the vertical axis plots the associated index
size in megabytes. Data points are taken at 25 gigabyte intervals between 50 and 200 gigabytes.}
  \label{fig:s3bd_index_sizes}
\end{figure}

As the size of the dataset grows, it is important for the utility information created by the Cloud Processing Server (the central Index) to be as small as possible, so as not to increase already large storage requirements. To demonstrate the space-efficiency of \system, we show the size of the central index as the size of the dataset increases (seen in Figure \ref{fig:s3bd_index_sizes}). 

The size of the index is shown to increase in a strictly linear fashion, with a linear regression analysis showing a strong positive correlation with a coefficient of $r=0.99$.  The central index is on average $\sim$0.27\% the size of the dataset, adding very little to storage requirements.  This small size can be attributed to \system's method for extracting only a small number of key phrases from each document.

\subsection{Evaluating Accuracy of \system}
\subsubsection{Benchmark Queries to Evaluate Accuracy}
We derived a set of benchmark queries based on the information present in the RFC dataset.  For testing accuracy, we consider two major categories of queries which a user may desire to search.  In the first category, we consider a user who already knows which document they are looking for, but may not remember where the document is located in their cloud system or may not want to look through a large number of files to find it.  Such queries are typically specific and only a small number of documents should directly pertain to them.  An accurate search system is expected to bring up these most desired documents first.

In the second category, we consider a user who wants to find all of the documents related to an idea. For instance, considering our motivational case, the law enforcement officer searching for similar crimes. Such queries would be broad with many possible related documents, and an accurate search system is expected to bring up the most relevant ones first.

\boxfigsecond{
Category 1 - Specific:\\
\texttt{IBM Research Report (IRR)} \\
\texttt{Licklider Transmission Protocol (LTP)} \\
\texttt{Multicast Listener Discovery Protocol (MLDP)} 

Category 2 - Broad:\\
\texttt{Internet Engineering (IE)} \\
\texttt{Transmission Control Protocol (TCP)}\\
\texttt{Cloud Computing (CC)}\\
\texttt{Encryption (EN)}
\caption{Queries used for evaluating relevance. Queries in Category 1 target a small set of specific, known documents within the collection, while queries in Category 2 target a broad set of documents, not necessarily known to the user. }
\label{fig:benchmarks}
}

\begin{figure}
  \centering
  \includegraphics[width=0.7\textwidth]{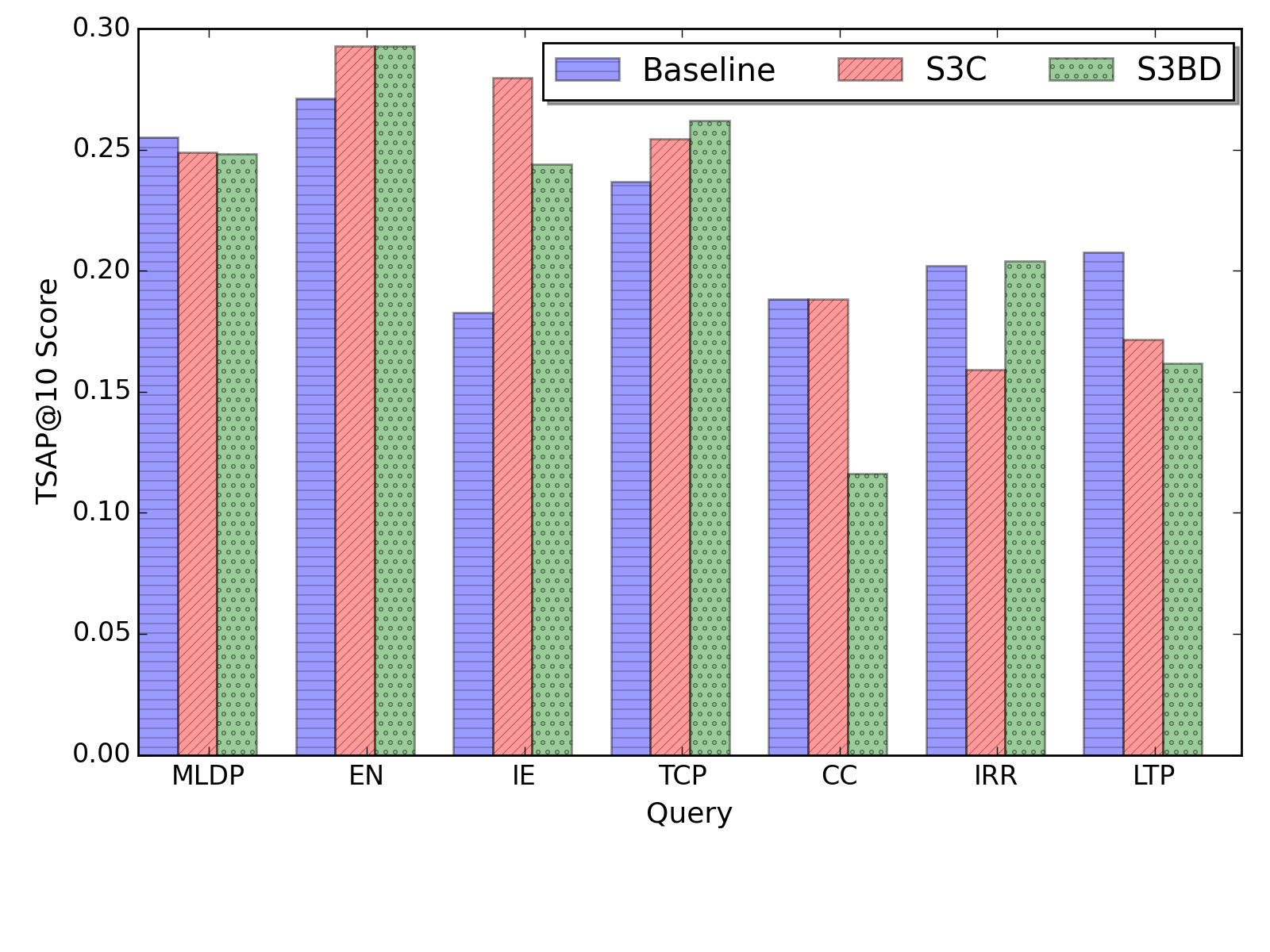}
  \caption{TSAP@10 score for specified queries for \system~, S3C, and a baseline system. For a given query, once the systems return a ranked list of results, a score is computed based on the human-determined relevance to each file.}
  \label{fig:relevance}
\end{figure}

\subsubsection{Metric for Evaluating Accuracy}
We define accuracy as how relevant the returned results are to the user's query, and how closely they meet user expectations. We describe accuracy in terms of the TREC-Style Average Precision (TSAP) method described by Mariappan \etal \cite{mariappan}. This method is a modification of the precision-recall method commonly used for judging text retrieval systems.  It is defined as follows:

\begin{equation}
Score = \frac{\sum_{i=0}^N r_i}{N} 
\end{equation}

Where $i$ is the rank of the document determined by the system and $N$ is the cutoff number (10 in our case, hence the term TSAP@10).  $r_i$ takes three different values:

\begin{itemize}
\item \texttt{$r_i = 1/i$ if the document is highly relevant}
\item \texttt{$r_i = 1/2i$ if the document is somewhat relevant}
\item \texttt{$r_i = 0$ if the document is irrelevant}
\end{itemize}

This allows for systems to be given a comparative score against other schemes in a relatively fast manner, without the need for knowledge of the entire dataset.

\subsubsection{Results of Evaluating Accuracy}
Figure \ref{fig:relevance} shows the TSAP scores (vertical axis) for different benchmark queries (abbreviated in the horizontal axis). For each benchmark query, we compare the relevance score of \system~compared to the scores of S3C and a baseline standard approach. In the baseline system, a simpler document representation is used (without keyword extraction) query modification is simpler, and there is no topic clustering, while the same Okapi search algorithm is used.

While \system~might intuitively seem to suffer in accuracy due to lower document representation within the shards, the figure expresses that is not the case. The relevance of the results obtained from \system, for most of them benchmark queries, are either the same or similar to those of S3C and the baseline. When compared to the less efficient baseline, S3BD provides better results for four of the benchmark queries. When compared to S3C, the relevance of S3BD is only lower in two benchmark queries, namely \texttt{Cloud Computing} and \texttt{Internet Engineering}. As a matter of fact, both of these benchmark queries are in the broad category. In the contrary, we observe that benchmark queries in the specific category have almost identical relevance in both systems.

We can infer that \system~provides higher accuracy with specific queries. The reason is that \system~can find shards for specific terms in the queries more accurately, in comparison to broad (\ie general) terms. In fact, for broad terms it is possible that pruning leads us to search less relevant shards. It is an interesting future research avenue to recognize broad terms and apply less aggressive pruning for them.


\section{Conclusion} \label{sec:conc}

In this research, we developed \system, a system to perform a secure semantic search over encrypted big data in the cloud.  \system~achieves real-time search ability on big data through \emph{pruning} irrelevant portions of the dataset at search time. \system~is comprised of three major architectural components, namely client application, cloud processing server, and cloud storage. After parsing and uploading documents, the cloud processing server clusters a central encrypted index into smaller, topic-based shards. At search time, client application compares the user's query to abstracted versions of those shards to determine the appropriate shards to be searched. \system~ontologically expands the user's search query to achieves semantic search ability. 

We performed analyses on \system 's performance and search accuracy using a working prototype. We analyzed the number of created shards to strike a trade-off between search performance and accuracy. 
Comparison of \system~ against similar works in the literature demonstrated that \system improves the search performance on the cloud by approximately 77\% without compromising accuracy.  

There are several avenues of research to extend \system. One interesting avenue is to determine the number of shards to be searched based on the broadness of the user's query. Another avenue is to dynamically determine the number of shards based on the dataset characteristics, \eg size. 
Adapting the \system~architecture based on the edge computing model can be explored to improve the efficacy of the search.


\section*{Availability}
Distributable .jars of the \system~core, as well as running instructions, are available at \url{http://hpcclab.org/products/S3BDJars.zip}. 

A preliminary version of S3BD with web interface for demonstration purposes is available at \url{https://teaching.cmix.louisiana.edu/~c00408440/S3C/S3Client/home.php}.

\section*{Acknowledgments}
We would like to acknowledge anonymous reviewers of the manuscript. 
This research was supported by the Louisiana Board of Regents under grant number LEQSF(2017-20)-RD-B-06, and Perceptive Intelligence, LLC.
Preliminary version of portions of this material were presented at the IEEE Big Data 2016 \cite{s3c}.

\footnotesize
\bibliographystyle{elsarticle-num} 
 \bibliography{references}

\balance
\end{document}